\documentclass[fleqn,twoside]{article}
\usepackage{espcrc2}
\usepackage{amssymb}

\usepackage{graphicx}
\usepackage[figuresright]{rotating}

\newcommand{\AmS}{{\protect\the\textfont2
  A\kern-.1667em\lower.5ex\hbox{M}\kern-.125emS}}
\newcommand{\gsim}{\;\raisebox{-0.9ex}
           {$\textstyle\stackrel{\textstyle >}{\sim}$}\;}

\title{ Photon 2003: a theorist's summary and outlook}

\author{Rohini. M. Godbole\address{Centre for Theoretical Studies,
        Indian Institute of Science,Bangalore, 560 012, India.}
             \thanks{Author wishes to acknowledge the hospitality of the 
          DESY-T division when part of this work was done}}
       
\begin{document}
\begin{titlepage}
\begin{flushright}
                                                   IISc-CTS/9/03\\
                                                   hep-ph/0311188  \\
\end{flushright}
                                                                                
\vspace{0.5cm}
\begin{center}
{\Large
                                                                                
{\bf Photon 2003: a theorist's summary and outlook  }}\\[5ex]
R.M. Godbole\footnote{e-mail:rohini@cts.iisc.ernet.in}\\[1.5ex]
                                                                                
{\it Centre for Theoretical Studies, Indian Institute of Science, Bangalore,
560 012, India.}
\end{center}
\vspace{1.0cm}
{\begin{center}
ABSTRACT
                                                                                
\vspace{2cm}
                                                                                
\parbox{15cm}{
In this talk I present a summary of some of the discussions at the
conference on various  topics in Photon physics, selected with a view to
give a theorist's perspective, of the current status and future prospects, of
the developments in the field. After discussing some of the recent theoretical
developments in the subject of Photon Structure function, I will focus on what
the photon has helped us learn about the spin structure of a proton, aspects
of perturbative and nonperturbative QCD as well as forward and diffractive
physics. I will discuss the challenges that the data on heavy flavour
production in the two photon  reactions and in photo production, seem to
have presented to the theorists.
Then I discuss the direction in which photon physics is likely to go in future
and what we {\em need} the photons to still tell us. I will end by talking
about the newer developments in prospects for photon studies at future
colliders and opportunities that these will provide us to learn about
the physics beyond the Standard Model.
}
\end{center}}

\vfill
{\begin{center}
{\it Summary talk presented  by R.M. Godbole at}                       \\
{\it PHOTON-2003, International Meeting on Structure and Interactions of the Photon} \\
{\it Frascati, Italy, April 7-11, 2003}
\end{center}}
                                                                                
\vfill
\end{titlepage}

\begin{abstract}
In this talk I present a summary of some of the discussions at the 
conference on various  topics in Photon physics, selected with a view to 
give a theorist's perspective, of the current status and future prospects, of
the developments in the field. After discussing some of the recent theoretical
developments in the subject of Photon Structure function, I will focus on what
the photon has helped us learn about the spin structure of a proton, aspects 
of perturbative and nonperturbative QCD as well as forward and diffractive 
physics. I will discuss the challenges that the data on heavy flavour 
production in the two photon  reactions and in photo production, seem to 
have presented to the theorists.  
Then I discuss the direction in which photon physics is likely to go in future
and what we {\em need} the photons to still tell us. I will end by talking 
about the newer developments in prospects for photon studies at future 
colliders and opportunities that these will provide us to learn about 
the physics beyond the Standard Model.
\vspace{1pc}
\end{abstract}

\maketitle

\section{Introduction}
In 2005 we will be celebrating 100 years of the discovery of the photon as 
the quantum  of electromagnetic field. During this period, photons have 
literally `illuminated' the path of important developments in our knowledge of 
fundamental physics. The beginnings of the quantum theory were in realizing
that radiation is quantized, photon being the quantum. It was the study of 
atomic spectra involving real photons that led us to a correct understanding 
of atomic structure.  The study of interaction between photons and electrons 
ushered in the era of Quantum Gauge Field Theories which now form the 
paradigm for the description of all the interactions among the fundamental 
constituents of matter, the quarks and leptons.  Probing the protons 
with virtual photons (DIS) led us to our understanding of the proton structure.
Photons have the amazing `self-analyzing' ability. As a result one has been 
able to study the structure of atoms, nuclei,proton and {\em photon} using 
photons. The last makes photon processes a particularly interesting laboratory 
to study  QCD dynamics. The `photon' conferences of which the current one 
is the 15$^{th}$, have had their own character and history. In the early 
years  mainly the photo production and $2 \gamma$ processes, the 
latter involving mainly spectroscopy, used to be the topic of discussion.
Since the past ten years these have also included discussions of 
 $F_2^\gamma$ studies in $2 \gamma$ processes at the $e^+ e^-$ colliders 
and study of `resolved photon' processes~\cite{Drees:1995wh} in hadronic 
interactions of high energy, real or `quasi-real' photons.  This means 
they have included, along with the discussion of physics at  $e^+e^-$ 
colliders TRISTAN and LEP, also a discussion of proton and photon structure 
as studied at the ep collider HERA.  
Thus these  photon conferences  have continued to reflect the use of 
photon physics as one big physics laboratory to study QED and QCD.

Highlights of what the community learned about the hadronic structure of the
$\gamma$ at this conference, have been described in the experimental
summary\cite{alex}.  I will include in the discussions below a theorist's  
perspective as to how this knowledge can (and has) improved our knowledge of 
$\gamma$ interactions. I will essentially discuss what the photon has taught 
us. After discussing the new theoretical developments  in the 
subject of  `real' photon Structure function, I will focus on what
the photon has helped us learn about the spin structure of a proton,
aspects of perturbative and nonperturbative QCD as well as diffractive physics.
I will discuss the challenges that the data on heavy flavour production in two 
photon  reactions and in photo production seem to have presented to the 
theorists. Then I will talk about the direction in which photon physics 
is likely to go in future and what we {\em need} photons to still tell us. 
I will end by talking about newer developments in the prospects at future 
colldiers for 
photon studies and opportunities these will provide us to learn about 
physics beyond the Standard Model. The choice of topics chosen reflects 
mainly a personal bias but also the theoretical developments and 
ideas concerning the topics covered in the experimental summary\cite{alex}.

\section{Structure function of the Photon and Proton: hard probes}
In this section I discuss a new PDF for real photon that was presented
at the conference, the issue of jet and heavy flavour production in
photon induced processes, news from the proton structure function at high 
$Q^2$ as well as the interplay between forward physics in DIS and structure of 
the virtual photon. I end by discussing  a prediction for beauty production 
in $\gamma \gamma$ collisions using  lessons learned from a study of 
$F_2^p$ at low-$x$ and the  information photons have provided on the 
spin structure of the proton
\subsection{Structure of Real Photon: New theoretical developments}
\label{21}
The basic terminology and concepts involved in the description of the 
`hadronic' structure of the photon have already been introduced\cite{alex}.
In spite of the experimental intricacies involved in extraction of 
$F_2^\gamma$ from the $2 \gamma $ processes in $e^+e^-$ experiments at LEP, the
data on $F_2^\gamma$ have come of age, even if the precision is  nowhere 
near the 
accuracies achieved in the $F_2^p$ measurements\cite{f2p}.  The available data
\cite{f2gamma} now cover a rather large range of $x$ and $Q^2$ and thus 
call for increasing  theoretical sophistication in obtaining a 
parametrization of the parton densities in the photon, the ones normally in 
use being the ones from the Dortmund group\cite{grvphot,grsphot}.

A new LO analysis of the structure of a `real', unpolarized  $\gamma$ 
was presented at this conference\cite{maria_tlk}.  They have 
used\cite{Cornet:2002iy} all the available high $Q^2$ data 
from LEP on $F_2^\gamma$ in the fit. But more importantly they have 
given a refined treatment of the heavy flavour in the LO parametrization. 
The mismatch between the theoretical prediction and experimental 
observations for  the $b \bar b $ production in $\gamma \gamma$ 
collisions~\cite{bbr2gm} makes the case of looking into this issue more 
carefully, even stronger. A parton of heavy flavour $h$ in any hadron, in this
case the $\gamma$,  can be treated in two different ways: one is the fixed
flavour number scheme (FFNS) where the light quarks are the only massless 
partons along with the gluons. The heavy quarks in the hadron are produced in
the Bethe Heitler process $G^\gamma \gamma^* \rightarrow h \bar h$. While this
scheme is the correct one at scales $\sim 4 m_h^2$, for higher scales a 
scheme  which treats even the massive  quarks to be `massless' partons and 
includes them in the evolution equation, the zero mass variable flavour number
scheme (ZVFNS), is the more appropriate, as it also resums the large 
logarithms. Of course results of the ZVFNS approach have to match 
smoothly with that of the FFNS one in the threshold region, $W \sim 2 m_h$,
where FFNS is the only correct description.
Hence, schemes combining both the approaches, where the number of the `active'
flavours in the hadron changes with the scale, the variable flavour number 
schemes, (VFNS's)  are the desired ones.  The ACOT group\cite{acot} has 
suggested, and implemented for the case of the proton, a kinematic solution to 
handle the threshold matching problem, the so called $ACOT (\chi)$ scheme.  
This involves, introducing a scale $\mu$ which decides the number of `active'
flavours in the  hadron, i.e., in the evolution equations and also a new 
variable $\chi_h \equiv  x(1+4m_h^2/Q^2)$, where $-Q^2$ is the  invariant mass
of the probing photon.

The new LO analysis~\cite{Cornet:2002iy} has implemented the ACOT($\chi$) 
scheme  for the $\gamma$. 
\begin{figure*}
\includegraphics*[scale=0.6]{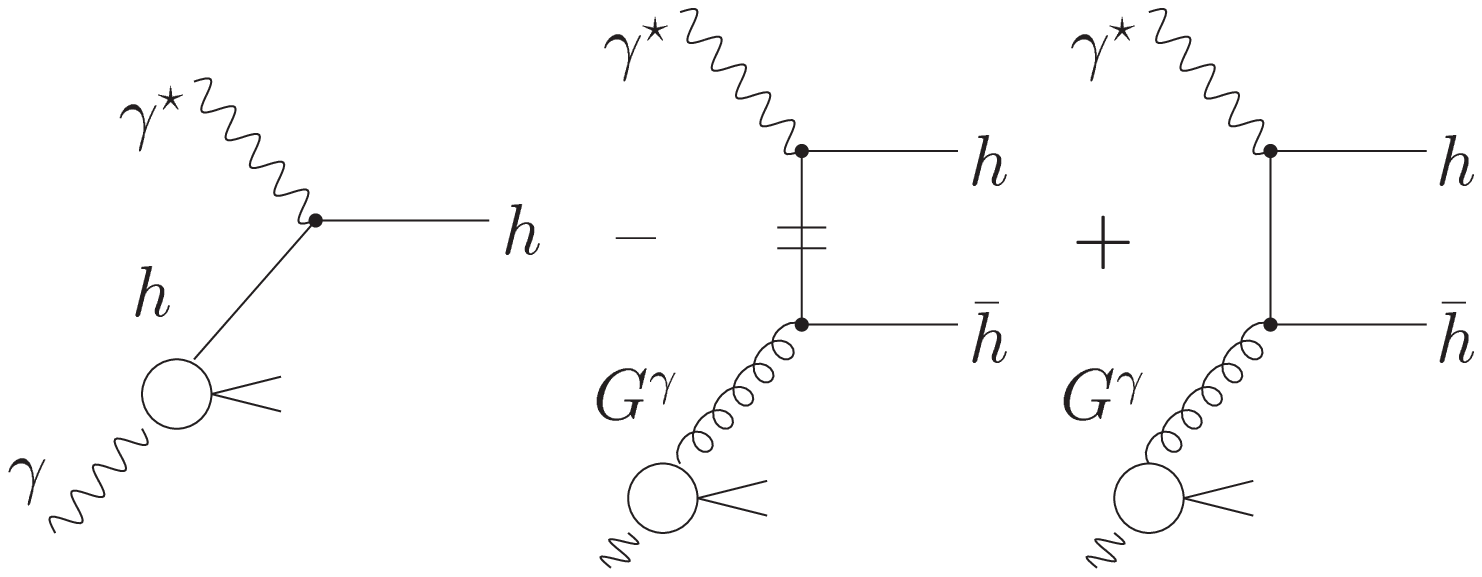}
\includegraphics*[scale=0.6]{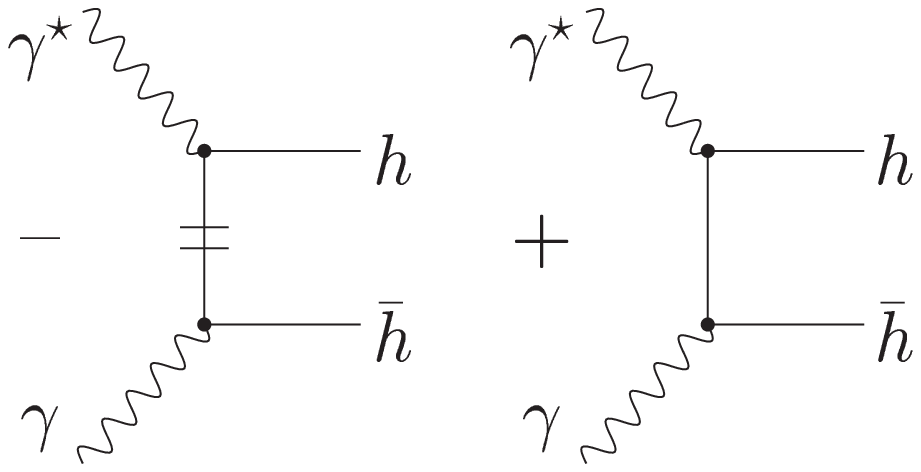}
\caption{Direct and resolved contributions to $F_2^\gamma$ in the ACOT ($\chi$) model.
The first diagram represents the ZVFNS contribution. The third (fifth) diagram
shows the FFNS contribution of the resolved (direct) photon, while the second
(fourth) diagram is the corresponding subtraction term\protect\cite{maria_tlk}.}
\label{fig1}
\end{figure*}
While treating the $h$ quark as a constituent in the photon
the contribution of the Bethe-Heitler diagram, where the $t$ channel $h$ quark
is on shell, is already included and hence has to be subtracted to avoid 
double counting. This model is diagrammatically indicated in Fig.~\ref{fig1}.
They compare it with a ${\rm FFNS}_{cjkl}$ fit obtained using FFNS.  The 
$\chi^2$ per degree of freedom for the ACOT$(\chi)$ scheme fit obtained 
by the authors, is somewhat better and the description of the large $Q^2$ 
and large $x$ OPAL data is also a little better than all the FFNS 
parameterizations including ${\rm FFNS}_{cjkl}$. CJKL use the same 
philosophy, of generating the parton densities radiatively, as that used 
in the GRV/GRS scheme. However, for the initial conditions, they follow 
the more reasonable, original GRV ansatz and use an incoherent mixture of the 
vector meson state, as opposed to the coherent one used in the GRS 
parametrization. This changes the relative importance of the $u$ quark 
densities to that of the $d$ quark densities, due to the 
difference in the charge of the $u$ versus $d$. As a result of this one has 
$ ( u/d )_{GRV} <  ( u/d )_{CJKL} < ( u/d )_{GRS}$. This can indeed be 
tested at the future $e \gamma$ colliders in a study of both the charged 
current and neutral current processes $e + \gamma \rightarrow e + X$ and
$e + \gamma \rightarrow \nu + X $, respectively, as pointed out in 
Photon-2001\cite{zerwas01}. An NLO analysis of the PDF is in progress.

\subsection{Jet Production in $\gamma \gamma$ collisions: a new NLO calculation}
The discrepancy between the L3 data on jets in $\gamma \gamma$ processes and 
the theoretical predictions\cite{Klasen:1997br} has already been 
discussed\cite{alex}.
Since the discrepancy exists only in case of L3 data and large $p_T^{jet}$, it
is unlikely that this would be due to our ignorance of the structure of the
`quasi-real' $\gamma$. Since it occurs at the edge of the phase space region, 
it could be due to higher order corrections. The existing NLO calculation
has been done by the slicing method. In a presentation at the 
conference\cite{bertora} the same NLO calculation done using the subtraction 
method was reported. Very schematically, in the slicing and subtraction 
method the NLO result can be written as, 
\begin{eqnarray}
\label{slice}
< F(x) >|_{slicing}^{NLO} &= \int_0^{1-\delta} dx {F(x) \over {(1-x)}} \nonumber \\ 
&+ F(1) Log (\delta) \\ [0.3cm]
\label{sub}
< F(x) >|_{subtraction}^{NLO} & = \int_0^1 {{F(x) - F(1) \theta(x -1 +x_c)} \over {(1-x)}} \nonumber \\ 
& + F(1) log (x_c)
\end{eqnarray}
In the slicing method, the phase space is divided into slices and  in the 
part of the phase space where the kinematics is collinear,  the contributions
involving virtual and soft real quanta, are canceled numerically against 
each other. It is clear from the two terms involving $\delta$ on the right 
hand side  of Eq.~\ref{slice} that the cancellations are large  as $\delta$ 
is small.  Also this involves dropping terms of ${\cal O } (\delta)$.  
In the subtraction method, $x_c$ is not necessarily small as the only 
condition on $x_c$ is $0 < x_c <1$. Hence the cancellation between the  soft 
real term and the virtual term is done without any approximations  and 
analytically.  A finite remainder is then integrated numerically  which does
not involve large numerical cancellations.

The results of this calculation agree well with those done by the slicing 
method before\cite{Klasen:1997br}. Their  results for inclusive distributions 
agree with OPAL data as well. The calculation also shows 
that even-though, in principle, the OPAL analysis of the dijet data with 
somewhat symmetric $E_T$ cuts could  have had IR problems, in practice  
the region chosen by OPAL is  free of the possible IR 
problems. In short two different NLO calculations of inclusive jet 
cross-sections in $\gamma \gamma$ collisions, agree with the OPAL data 
but have trouble with the L3 data, thus deepening the mystery.

\subsection{Charm content of the photon and heavy flavour production in 
real photon induced processes} 
In general, definition of the heavy flavour content of any  hadron has
theoretical ambiguities (cf. discussion in section~\ref{21}) and the issues
are somewhat more complicated for $\gamma$ due to the blurring of the boundary 
between the `resolved' and `direct' processes beyond the LO. It had been 
pointed out earlier\cite{Drees:1995nu}  that inclusive photo production of the 
$D^*$  could be used as a `direct' probe of the charm content of the photon.

\begin{figure}[htb]
\includegraphics*[scale=0.5]{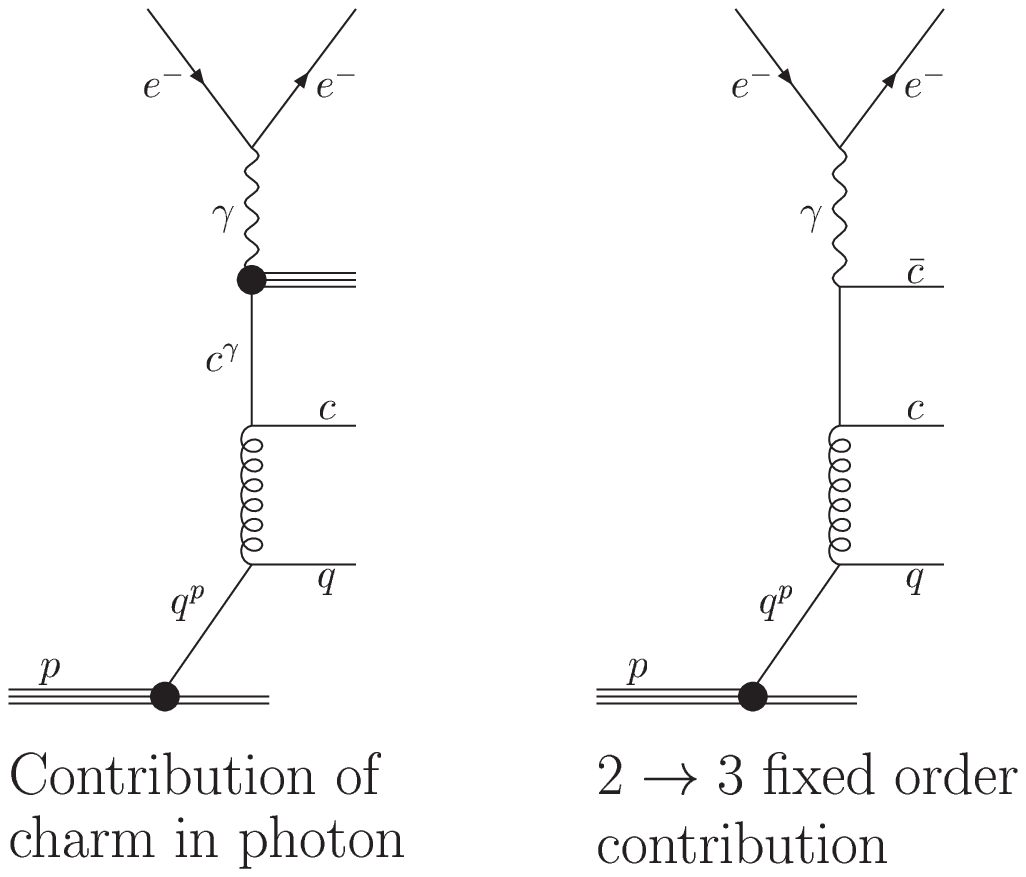}
\caption{Diagrams showing `resolved' contribution to inclusive production 
of $D^*$ with a jet as charm excitation on the left or as 
contribution from a fixed order (FO)  $2 \rightarrow 3$ process on the 
right.}
\label{fig2}
\end{figure}

In the LO the `resolved' contribution to the photo production of a $D^*$ along 
with a jet, comes  from diagram similar to that shown on the left in 
Fig.~\ref{fig2}.  This can 
be looked upon as `excitation' of the charm in  $\gamma$. This can also be seen
as a part  of the contribution of the `fixed order' (FO) 
$2 \rightarrow 3$ processes, when the $c$ quark in the $t$ channel 
in the diagram shown in the right panel of the same figure is 
almost on mass shell.  Actually  a LO calculation in the former 
language might be better than just a simple FO calculation, as the former will 
sum up the leading large $p_T$ logs. As a matter of fact, our 
calculation\cite{Drees:1995nu} had also demonstrated that one could enhance the
`excitation' contribution in  the sample of the inclusive $D^*$ events, by 
making cuts on the rapidity of the away side jet, to eliminate contribution 
coming from the `direct' process  $\gamma + G^p \rightarrow c + \bar c$. 

A very nice analysis of the charm dijet photo production\cite{ukarshon,zeuschrm}
presented at the conference, shows a clear evidence for events originating
from the charm in the photon. 
\begin{figure}[htb]
\includegraphics*[width=17pc]{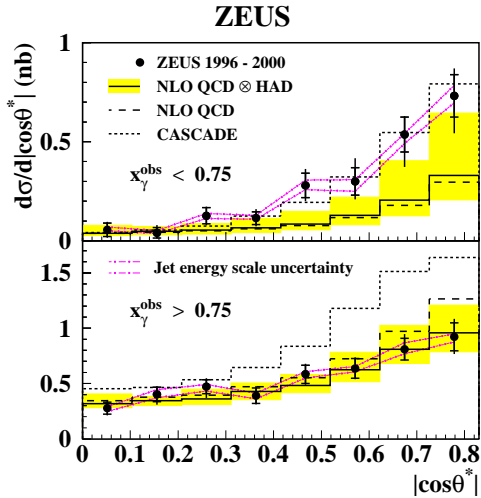}
\caption{Charm dijet angular distribution in photo production, $\theta^*$ being 
the angle between the jet-jet axis and the beam axis in the dijet rest frame\protect\cite{ukarshon}.}
\label{fig3}
\end{figure}
For the dijets arising from $\gamma + G^p \rightarrow c + \bar c$, the 
hard subprocess involves exchange of a $c$ quark in the $t$ channel whereas
for the dijets coming from the hard subprocess shown in Fig.~\ref{fig2}
it is $g$ which is exchanged in  the $t$ channel. 
The spin 0  and spin 1 character of the exchanged quanta would imply an angular
distribution $\propto (1 - |cos \theta^*|)^{-1}$ and $\propto 
(1 - |cos \theta^*|)^{-2}$ respectively.  The observable 
$$
x_\gamma^{OBS} = {{\sum_{jets} E_T e^{-\eta}} \over  {2 y E_e}}
$$
gives the fraction of the photon momentum entering the subprocess producing 
the dijet.  Thus one expects $x_\gamma^{OBS} = 1$ for the direct process and
$ x_\gamma^{OBS} < 1$ for the resolved process. A cut of $0.75$ on this 
variable thus separates the `direct' and `resolved' enriched part of the
event sample.  The upper and lower panels of Fig.~\ref{fig3}, show the
angular distributions for $x_{\gamma}^{OBS} < 0.75$ and 
$x_{\gamma}^{OBS} > 0.75$ respectively.  It is clear that the `resolved' and
`direct' samples are consistent with the $g$ and $q$ exchange respectively, 
thus showing clear evidence for the resolved photon charm excitation pointed 
out earlier~\cite{Drees:1995nu}. The figure also shows that this conclusion is
pretty robust and unaffected by the hadronisation model. About $40\%$ 
contribution comes from the `resolved' photons and hence mainly from the 
$c$ in the $\gamma$. 

The observed distributions agree well in shape with a LO  Monte Carlo. 
To understand the comparisons of the data with NLO calculations, one has
to discuss different schemes that exist for doing these, the differences being 
due to two different ways of looking at heavy quark content of a hadron,
mentioned already in section~\ref{21}.
The massive FO NLO calculation\cite{Frixione:2002zv}  contains, in principle, 
both the `direct' and `resolved' contributions, the scheme being valid only for 
$p_T, Q \sim m_c$.  Further in FFNS ($m_c \neq 0$ and three active flavours 
in $p$ and $\gamma$) the `resolved' 
contribution is effectively treated only in the LO and hence has a 
large scale uncertainty. Further fragmentation function again is only 
perturbative and does not include evolution. This calculation, however, does 
include the {$m_c^2/ p_T^2$} terms correctly.  For $p_T >> m_c$  
large logs need to be resummed.  Calculations can also be performed in 
ZVFNNS  where 
the $c$ is treated as an active flavour in the $p, \gamma$, the scheme 
being valid only at scales $Q,p_T$  all  much larger than 
$m_c$\cite{Kniehl:1996we}. Of course in this case the $2 \rightarrow 3$ 
contribution needs to be subtracted, to avoid double counting, just as in 
the case of $F_2^\gamma$ discussed in section~\ref{21}.  Further, one 
needs also to match the FO massive and massless calculations. In the 
matched FONLL calculation~\cite{Cacciari:2001td}, ${m_c^2/p_T^2}$ 
mass effects are incorporated up to NLO and resummation of the $p_T$ logs is
done up to the NLL level. At this conference the massive  variable flavour 
number scheme, massive-VFN, where one uses $m_c \neq 0$ and still $c$ is an
active flavour in the initial state, was discussed~\cite{kramer_f} in 
the context of charm production in $\gamma \gamma$ collisions. This scheme has
the advantage that it includes both, the ${m_c^2/ p_T^2}$ effects 
due to the nonzero mass of $c$ and the $\log {p_T^2/ m_c^2}$ effects in
the evolution and the fragmentation function of the initial and final state 
charm, at large $p_T^2$, to NLO. At low $p_T$ however, in the limit of
$m_c = 0$, the massive VFN scheme does not reduce to FFNS.
The two differ by finite terms, which  have been calculated
\cite{kramer_f} for the direct and once resolved terms in $\gamma \gamma$
production of heavy flavour. The matching of the massive, resummed calculation
to FFNS, at small $p_T$ still needs work and is in progress. 

To return to the data on the photo production of $D^*$ with a jet, 
\begin{figure}[htb]
\includegraphics*[width=17pc]{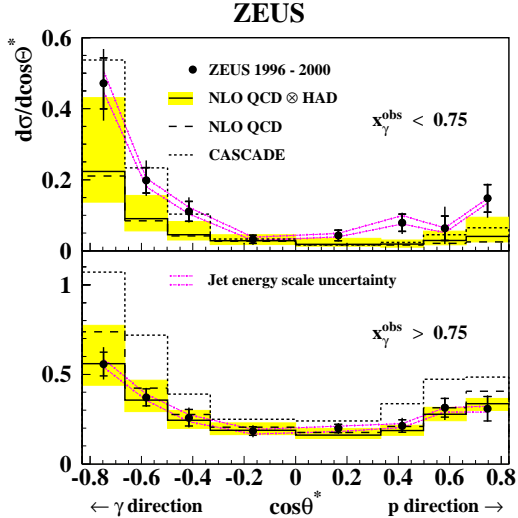}
\caption{Comparison of the data with a FO NLO calculation with/out
hadronisation effects as well as with CASCADE Monte Carlo, for the `direct'
and `resolved' enriched sample with $x_{\gamma}^{OBS} > 0.75 (< 0.75)$\protect\cite{ukarshon}.}
\label{fig4}
\end{figure}
only NLO calculations available in the public domain are the FO, 
FFN\cite{Frixione:2002zv} and FONLL\cite{Cacciari:2001td} and 
have been used to make comparisons with the data. 
Comparison of the FO, FFNS calculation, including hadronisation effects, shown
in Fig.~\ref{fig4}  indicates that for $x_{\gamma}^{OBS} > 0.75 $ the NLO 
calculation describes the data well. However, for the lower $x_{\gamma}^{OBS}$ 
the NLO calculations fall short of the data in both the $p$ and $\gamma$ 
direction. CASCADE, a Monte Carlo which implements CCFM evolution, 
overestimates the $x_{\gamma}^{OBS} > 0.75$ contribution.
The underestimation of the low $x_{\gamma}^{OBS}$ tail of the data by the FO,
NLO prediction is confirmed further by comparisons of more differential 
distributions, made possible by large data sample, in $\eta (D^*)$ for 
different bins of $p_T(D^*)$. The data are either close to or above the upper 
band of FO,FONLL predictions for medium $p_T (D^*)$ and forward $\eta (D^*)$.

Given the large size and precise nature of the data sample, a comparison of 
these data with a  massive VFN calculation where the resolved contribution is 
also treated to NLO, is indeed the order of the day.  These data  
provide a very nice  laboratory to learn how to implement the heavy quark in
the initial as well as in the final state. The fact that at LEP, 
data on  $D^*$  production in $\gamma \gamma$ processesfrom all the three LEP
collaborations, ALEPH, OPAL and DELPHI, for $p_T(D^*) > 2 $ GeV,
is explained very nicely in terms of a massive, resummed NLO 
calculation\cite{kramer_f}, makes similar comparison for the photo production 
data quite imperative.  Since the `away' side jet will be a light 
quark jet for the `resolved' contribution and will contain a $D^*$ 
for the direct contribution, one could also use this for separating the two.

\subsection{Proton Structure Function: news from large $Q^2$ region}
In the six years of running, two experiments at HERA have collected an
impressive amount of data on $F_2^p$ over  a wide range
$1 < Q^2 < 4 \times 10^4 {\rm GeV}^{2}$ and
$6.2 \times 10^{-5} < x < 0.65 $, the coverage at low-$x$ being
restricted to smaller $Q^2$ values and that at large $x$ to
$Q^2 \gsim 2.5 {\rm GeV}^2$ as  reported at this
conference\cite{f2p}. The PDF's  obtained
using QCD fits to HERA data alone, show very nice consistency with those
obtained from a global analysis. The errors on  the $u (d)$ quark densities
in the proton obtained from the HERA data are between $1$--$2\%$ ($3$--$10\%$)
over the $x$ range $0.01 < x < 0.4$, whereas the errors on the quark sea and
gluon densities, which dominate the $F_2^p$ at low-$x$, are $\sim 5 \%$,
and $10 \% $ for $x < 0.1$. Of course this is great news from the point of
view of making more accurate predictions for the LHC possible.

More interesting from a theorist's point of view, is the fact that the
errors in the reported values of  $F_3$ from the measurements of cross-sections
for the charged current process $e + p \rightarrow \nu + n$, $\sigma_{CC}$,
which are statistics dominated, will reduce substantially after HERA-II running
and hence can be used for some interesting tests of QCD.  HERA-II low
energy running has already yielded measurement of $F_L$. The reduced
cross-sections ${\tilde \sigma}_{CC} $, where the factors involving $Q^2$,
propagator and the coupling of the leptons to the $W$ are removed, for incident
$e^-$ and $e^+$, can be simply written in terms of the Quark Parton Model
expressions involving the known PDF's.
\begin{figure}[htb]
\includegraphics*[scale=.40]{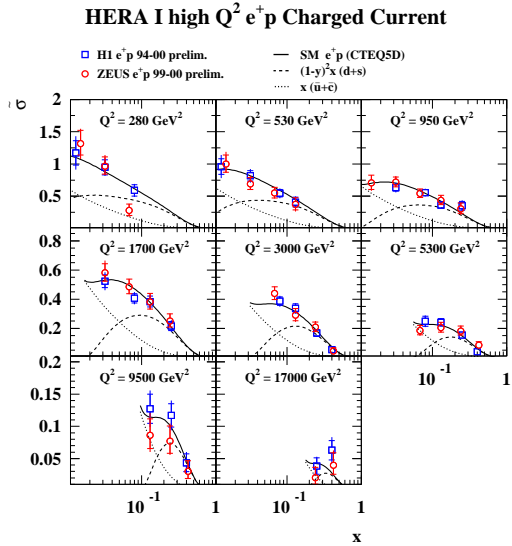}
\caption{ Data on ${\tilde \sigma}_{CC} $ at high $Q^2$ confronted with
predictions  from the PDF's obtained with the $F_2$ fits\protect\cite{f2p}.}
\label{fig7}
\end{figure}
Fig.~\ref{fig7} shows measurements of ${\tilde \sigma}_{CC} $,
for $280 < Q^2 < 17,000$ GeV$^2$ covering two orders
of magnitude in $x$, against the predictions given by PDF's extracted
from the data on $F_2^p$. The consistency also indicates that there are no
special surprises in the large $Q^2$ region covered up to now.

\subsection{DIS and hadronic structure of virtual photon}
HERA has provided interesting information on hadronic structure of the 
`quasi' real photon as well as the virtual photon, via inclusive jet production
and heavy flavour production. The new experimental results in this respect 
were covered in the experimental summary\cite{alex}. As far as the proton is 
concerned, low-$x$ is the second frontier of interest and investigations
at HERA. Some of the new results in the jet production in the `forward region'
in the DIS, where BFKL dynamics for the proton is likely to leave its imprint, 
indicate that the hadronic structure of the virtual photon might be playing an
important role in the forward region. This also underscores the need 
to understand the structure  of a virtual photon, if one wants to develop
clean diagnostics to study the proton dynamics at low-$x$. Forward physics also
holds the key to an understanding of the QCD dynamics of hadronisation. In this
context the data on forward particle and jet production in DIS presented at 
this conference\cite{jaceck} have thrown open yet another challenge to the 
theorists. 

Inclusive jet data in DIS were analyzed with a special focus on the forward 
jets in the H1 data, for $5 < Q^2 < 100~{\rm GeV}^2$ and $0.2 < y < 0.6$. The 
description of the data by a DGLAP, NLO calculation slowly deteriorates as one 
goes from backward to forward direction\cite{jaceck}. The deviations are
particularly large for small $Q^2$ and small transverse energy $E_T$. 
\begin{figure}[htb]
\includegraphics*[width=17pc]{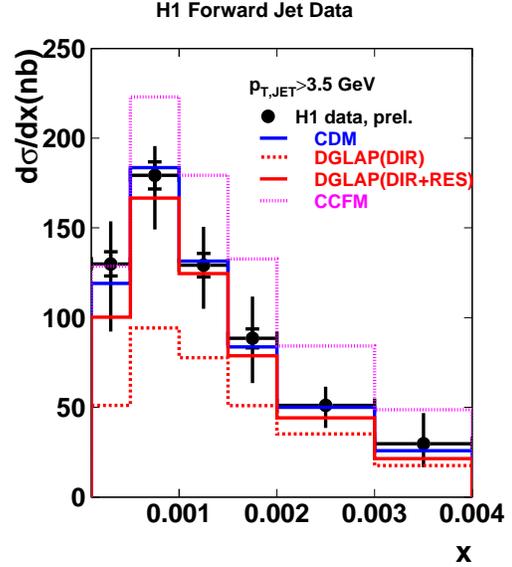}
\caption{Comparison of the H1 forward jet data (1997) with DGLAP LO and
CCFM predictions, for  low $x$\protect\cite{jaceck}.}
\label{fig5}
\end{figure}
The comparison between the data on forward jets (with cuts to enrich the 
possible BFKL contribution) and calculations, shown in Fig.~\ref{fig5},
makes it clear that the data can be understood in terms of 
a LO DGLAP calculation which includes both the `direct' and `resolved' 
contribution. However, one has to choose a scale $Q^2 + p_T^2$, for the
structure of a virtual photon with virtuality $Q^2$; the $p_T^2$ values being
$0.5 < {p^2_T/ Q^2} < 2$.  An earlier theoretical analysis which treated 
the `direct' contribution to NLO and included the virtual contribution as 
well\cite{Kramer:1999jr}, also needed to choose a similar scale. The CCFM 
model which should include effects of the BFKL dynamics at low $x$ overshoots 
the data.

It is expected that features specific to BFKL dynamics might be washed out 
due to averaging if one looks at inclusive quantities at HERA. But study of 
exclusive  variables chosen to maximize the effect of the BFKL dynamics 
whould offer a good probe.
Thus forward particles are expected to be  a better probe of the BFKL 
dynamics than forward jets. 
\begin{figure}[htb]
\includegraphics*[width=17pc]{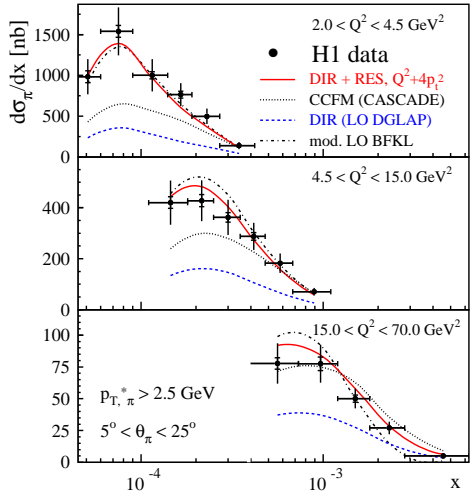}
\caption{Comparison of the H1 forward particle data  with DGLAP LO and
CCFM predictions\protect\cite{jaceck}.}
\label{fig6}
\end{figure}
The analysis is more difficult but results are now available and  were
presented at the conference\cite{jaceck}.  Comparison between the data 
and the DGLAP as well as CCFM calculations, shown in Fig.~\ref{fig6}, indicates
again that the {\it resolved} contribution {\it has} to be included for the 
best description of the data , but with  $Q^2 + 4 p_T^2$ as the scale for the
PDF in $\gamma$. 
A  modified BFKL calculation\cite{kwiecinski} seems to describe the 
data as well.  Clearly a complete NLO calculation, for the resolved as well 
as the  direct contributions to the forward particle production, in the usual 
DGLAP formulation is the order of the day. NLO calculation of the direct 
contribution are underway\cite{spanish,MF}. These seem to show trends similar 
to those seen in comparison of the data and FO,NLO calculation  of forward 
jets, viz. the resolved contribution is a significant part of the FO, NLO
corrections to the `direct' process. A complete calculation should enable 
us to understand virtual photon structure better and also to address 
whether these data hold any indication of BFKL dynamics.

Recall also the indications that the ZEUS data on jet production in 
DIS\cite{bussey} does not seem to be explained by  just a FO, NLO 
calculation of the `direct' process\cite{Catani:1996vz} in spite of the  
large $Q^2$ of the $\gamma$ involved and one might need 
to inlcude the resolved contribution to the NLO\cite{Potter:1998jt}, 
to explain the data in terms of a QCD calculation. Thus, the virtual photon
has indeed thrown up some new  challenges and these indicate more work for 
theorists.

\subsection{ Structure of the proton and heavy flavour production in
$\gamma \gamma$ collisions}
One of the important experimental presentations at the meeting
was the fact that while data on charm production in $\gamma \gamma$ collisions
from {\em all}  the four LEP collaborations are in agreement with the NLO 
theoretical calculations {\it all} the LEP experiments find too much 'beauty' 
in $\gamma \gamma$ collisions~\cite{alex,bbr2gm,kapusta} compared to 
the theoretical expectations. The observation is all the more puzzling since
the theoretical uncertainties coming from the QCD higher order corrections
and/or from the  photonic parton densities are expected to be smaller for the
$b \bar b$ than for $c \bar c$. Hence any new theoretical calculation of the
former is of interest. One such attempt is described below.

It is possible that the collective phenomena like gluon recombination or
saturation might play an important role because of the small values of 
$x$ being explored at HERA. In that case CCFM evolution might be the 
more appropriate one to be used \cite{smlxrev} for analyzing the data
on $F_2^p$.  The HERA data for 
$x < 5 \times 10^{-3}$ and $Q^2 > 4.5$ GeV$^2$, have been used\cite{jung} to 
determine the unintegrated gluon density in the proton, which  is the correct
one to use while calculating various quantities in the $k_t$ factorization
scheme rather than the usual collinear, DGLAP formulation.

In this approach the unintegrated gluon densities in  a hadron are given in 
terms of the parton densities at a small scale $Q_0$ by the evolution equation
governed by the CCFM splitting function. The scale $Q_0$ and the 
cutoff value of internal $k_T$, viz., $k_T^{cut}$ below which nonperturbative
region is entered, are the two parameters which are fitted to the data. Using 
\begin{figure}[htb]
\includegraphics*[width=15pc]{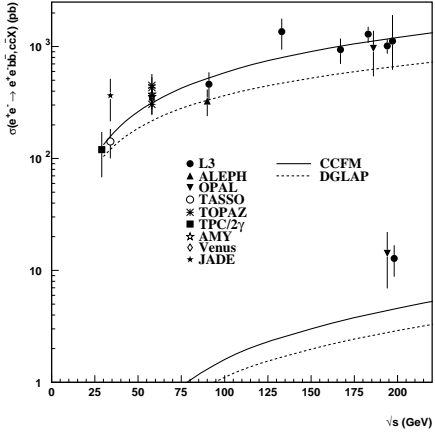}
\caption{Predictions for the heavy flavour production in $\gamma \gamma$
collisions in the $k_T$ factorization scheme\protect\cite{jung}.}
\label{fig8}
\end{figure}
the parameter values so determined\cite{jung}, one can
determine the unintegrated gluon densities in the case of a $\gamma$  as well.
In principle, these will include terms in higher order corrections 
beyond those included in the usual collinear approximation in the DGLAP 
formulation. Fig.~\ref{fig8} shows that while the charm production is 
reasonably well described. the beauty production, though somewhat higher than
the one obtained in the collinear approximation at NLO, is still much below 
the data. This calculation thus shows that the discrepancy between the data 
and theoretical prediction are unlikely to be due to higher order effects
neglected in the collinear, DGLAP calculation.

\subsection{Spin Structure of proton}
Most of the information on the spin structure of the proton normally 
has come from the polarized DIS.  The measurements of polarization
asymmetries of the inclusive DIS cross-sections can be used to extract the
helicity distribution $g_1$ in a proton which is the longitudinal spin 
distribution of the quarks in a longitudinally polarized proton. The function
 $g_1(x) $ measures the difference between the quark density 
distributions ($\Delta q$)  with its spin parallel and anti-parallel 
to the direction of spin for the longitudinally polarized photon, 
in the infinite momentum frame.  Once one moves on to semi-inclusive
DIS (SIDIS) and two hadrons are involved, then apart from the unpolarized 
structure function $f_1(x)$ and $g_1 (x)$, the only other twist 2 distribution 
function characterizing the spin structure of the proton, is the 
transversity distribution, which measures the transverse spin distribution 
in a target polarized transversely to the virtual photon\cite{Mulders:1995dh}. 
Since rotation and boost do not commute, $g_1 (x)$ and $h_1(x)$, are 
two distinct functions. As a matter of
fact the QCD evolution of the transversity and helicity distributions
are different. Also the axial and the tensor charge of the nucleon i.e.
the integral of the helicity and transversity distributions respectively are
different from each other and their values  as evaluated on the Lattice 
are 0.18 and 0.56\cite{Aoki:1996pi} respectively. Here the axial charge is 
the integral of the isoscalar combinations of $\Delta q$.

Currently the focus of studies in the spin structure of the proton at Hermes
and Compass,  is the transversity distribution $h_1(x)$ in the SIDIS. 
Since it is a chiral odd function and all the gauge 
interactions preserve helicity, a measurement of $h_1(x)$ is possible only 
when combined with chirally odd fragmentation function. These will give 
rise to single-spin azimuthal asymmetries in SIDIS.  There exist several 
different effects which could give rise to such chirally odd frgamentation
and has been modelled theoretically. The analyzing power of the single-spin 
azimuthal asymmetry can get diluted to some extent due to the higher twist 
effects, in case of targets which are longitudinally polarized. It is clear 
that these studies are of immense theoretical interest.

An analysis of the single-spin azimuthal asymmetries  ${\cal A}_{UL} (\phi) $ 
from Hermes, with unpolarized lepton but longitudinally polarized targets,  
was presented at the conference\cite{hash}. The data from the current run, 
show nonzero target-spin azimuthal asymmetry and these are consistent with 
predictions of models which are transversity related (at least for the pions). 
Measurements of the asymmetries with transversely polarized target, in 
Hermes-II, will provide  much better information on the transversity 
distribution. Beam-spin
azimuthal asymmetries as are being measured by the CLAs collaboration 
can probe the  Collins fragmentation function in combination with different 
distribution functions than one can do with polarized target. The results from 
Hermes-II  and CLAs hold thus a great promise  for spin physics.

\section{Structure of $p,\gamma$: Soft  and Diffractive Physics}
A variety of theoretical and experimental discussions were presented at the
conference on this subject. I choose to highlight the new results from HERA in 
diffraction which, along with the data from the Tevatron, have sparked a very 
intense theoretical activity in the area. These measurements have also 
implications for the energy dependence of the total cross-sections for the 
proton and photon induced processes and thus hold a lot of promise
for us to learn about the dynamics of QCD in the nonpeturbative regime. 
The exclusive, diffractive production of vector mesons has seen developments 
in the theoretical models recently which will, however, not be discussed for
reasons of stpace.

\subsection{Diffraction}
A summary of the  vast amount of data on hard diffraction at 
HERA\cite{hera-diff} as well as the information on diffraction from 
the Tevatron\cite{cdf-diff} was presented at the meeting.  The 
cross-section for the reaction $e p \rightarrow e p X$, can be written in terms
of the diffractive structure function $F_2^{D(4)} (\beta,x_{\mathbb P},t,Q^2)$.
The earlier HERA data had supported the Regge factorization hypothesis 
according to which one has,
\begin{eqnarray}
\label{regge}
F_2^{D(4)} (\beta, x_{\mathbb P},t,Q^2) &= f_{{\mathbb P}/p} (x_{\mathbb P} , t) F_2^{\mathbb P} (Q^2, \beta)
\nonumber \\ 
&= {e^{bt} \over {x_{\mathbb P}^{(2 \alpha_{\mathbb P} (t) -1)}}} 
F_2^{\mathbb P} (Q^2, \beta) 
\end{eqnarray}
where $x_{\mathbb P}, \beta$ are related to Bjorken $x$ by 
$x = x_{\mathbb P} \beta$.
New results from H1 and ZEUS for diffractive jet production at HERA were
presented at the conference and these covered a wide range of $Q^2, \beta$
and $x_{\mathbb P}$. The selection of diffractive events was done in more than
one ways.  $Q^2$ evolution of $F_2^{\mathbb P} (Q^2, \beta)$ in Eq. \ref{regge} 
is given
by the DGLAP equation. The data show scaling violation up to large values of
$\beta$, which in turn means large gluon component in the colourless exchange
which causes the diffractive scattering. This is indeed consistent with our QCD 
picture of this colourless exchange (i.e. the Pomeron) as a multigluon ladder. 
The data allow an extraction of the pomeron intercept $\alpha_{\mathbb P} (0)$
as a function of $Q^2$. Most interestingly the value of $\alpha_{\mathbb P} (0)$
is substantially higher than the 1.09 expected for the universal soft Pomeron,
even-though the errors on $\alpha_{\mathbb P} (0)$ are quite 
large\cite{hera-diff}.

Further, H1 has been able to make an NLO QCD fit to the data on Diffractive
structure function and they presented\cite{hera-diff} a set of 
diffractive parton densities. The ideas of QCD 
factorization will imply that the diffractive cross-sections for different 
hard processes be given by convolution of the diffractive parton densities 
with the partonic subprocess cross-section. This seems to work between the 
diffractive jet and charm production at HERA. However, the factorization 
breaks down when one uses the same densities to calculate the diffractive
jet cross-section measured at the Tevatron\cite{cdf-diff}, where the 
diffractive events are selected by requiring rapidity gaps. This confirms the 
breakdown of factorization between the $ep$ and $\bar p p$ case, noticed 
before. This strong 
violation of factorization can be understood in terms of the additional 
spectators present in the $\bar p p$ environment. 
However, a recent analysis \cite{bloistalk} by H1, presented after the 
conference, of the diffractive charm production in resolved photon processes, 
puts some doubt on this understanding of  violation  of factorization. 
Therefore, hard diffraction is  certainly an area to look to-wards when 
HERA-II data start coming out. Theory interest is not just due to the 
insight in the nonperturbative /semi-hard QCD dynamics that one expects these
data to provide, but also from a very pragmatic point of view of being able
to calculate cross-sections of different diffractive processes, at the 
Tevatron and the LHC, where these might give novel search channel for the Higgs.

\subsection{Total cross-sections}
\label{32}
In the Regge-Pomeron theory one expects the rise with energy of total 
cross-section to be given by $s^{\alpha_{\mathbb P}(0) -1}$. One possible
implication of a value of $\alpha_{\mathbb P}(0)$ substantially larger than the
universal soft pomeron value of $1.09$ as  reported above, albeit with large 
errors, is that the cross-sections might rise faster with energy than the
expected universal $s^{0.09}$ behavior. The high energy behavior of the total
cross-sections was a subject of much discussion at the conference. No new 
results  were presented on the experimental side, though new analyses 
from H1 and OPAL are expected to yield new numbers for the total 
$\gamma p$ and $ \gamma \gamma$ cross-section soon. The extraction of the 
total hadronic cross-sections involving photons from the measurements of
the $ep$ and $e^+ e^-$ processes, is  beset with the same problems as in 
extraction of $F_2^\gamma$, explained in detail\cite{alex} in that context.

One of the  analysis\cite{block} presented  studies all the cross-sections in
a global picture and a  fit is made in a QCD inspired model, 
using unitarity and factorization.  An eikonal picture is used.
Normally the  eikonal is  determined  by the parton densities in the hadron
and the QCD parton scattering cross-sections. In this model the coefficient of
the subprocess cross-sections appearing in the eikonal are not the product of 
the parton densities but are free functions which are fitted using the proton
data. Using these fitted functions and ideas of VDM and quark parton model, the
$\gamma p$ and $\gamma \gamma$ cross-sections are  obtained. The authors find
a good description of $\gamma p$ and $\gamma \gamma$ data with these fitted
parameters, if they use the L3, OPAL  data obtained by using  PHOJET for 
unfolding and {\em further renormalise both} the data downward by about 
$10 \%$. The global 
fit is much worse and requires a downward normalization of the data by about 
$20\%$  if the data obtained using  PYTHIA for unfolding are used. The 
\begin{figure}[htb]
\includegraphics*[width=17pc]{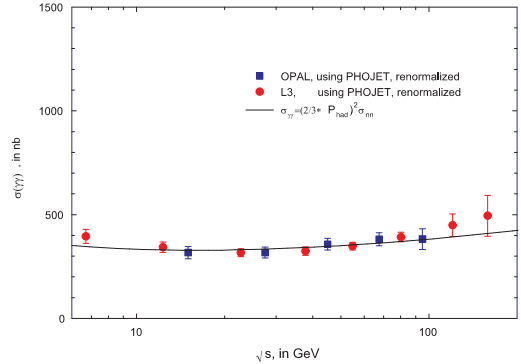}
\caption{Renormalized  OPAL and L3  data on $\gamma \gamma$ cross-sections,
given using PHOJET for unfolding, compared with global fit of \protect\cite{block}.}
\label{fig9}
\end{figure}
authors conclude from this that all the cross-sections rise universally. 
However, the L3 data, the only one data set to have points at higher energy, 
still systematically lies above their fit even after normalizing it downward
as can be seen in Fig.~\ref{fig9}.

The other sets of talk\cite{yogi,lia} on total cross-sections presented 
theoretical motivation of an unitarised eikonal model based on QCD where 
the  eikonal is calculated in terms of minijet cross-sections which can be
computed using perturbative QCD subprocess cross-sections 
$\sigma (p_1 p_2 \rightarrow p_3 p_4)$ and parton densities in the proton 
and photon as measured experimentally. In this formalism one has to make 
a model for the transverse overlap function of the two hadrons. One talk
\cite{yogi} outlined the basic idea of this model and the other one \cite{lia}
discussed the results obtained in the model for $pp/ p \bar p, \gamma p$ and 
$\gamma \gamma$ total cross-sections. The soft parameters of the model are 
fixed by fitting the $p \bar p$ and $p p$ cross-sections. This model gives 
a nice description of the initial {\em fall} and the subsequent {\em rise}. 
\begin{figure}[htb]
\includegraphics*[width=17pc]{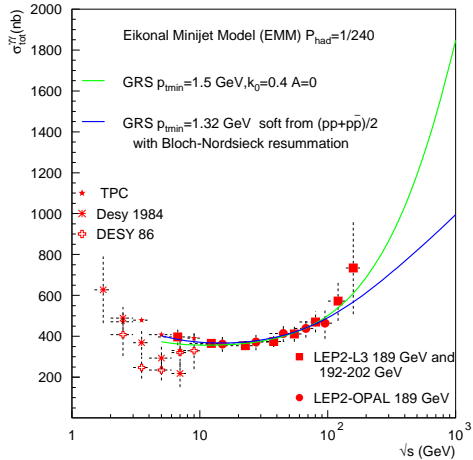}
\caption{All the data on the hadronic  $\gamma \gamma$ cross-sections,
compared with predictions of the Eikonalized Minijet Model and
the BN model\protect\cite{lia}.}
\label{fig10}
\end{figure}
Fig.~\ref{fig10} shows the comparison of the model predictions with data for
the $\gamma \gamma$ case, with the soft parameters in the eikonal being taken 
to be average of those fitted in case of $pp$ and $\bar p p$. The model gives
a good description of the observed energy dependence of the $\gamma p$ 
cross-section as well, provided the soft part in the eikonal is taken from only
the $pp$ fits. Thus here one seems to have a breakdown of exact factorization.
Also the BN form of the transverse overlap function of the two hadrons $A(b,s)$
is derived only in the LO and in Abelian approximation. Further, prediction for the elastic cross-section is off by about $10 \%$ in this approach.

Clearly the situation needs  clearing up. The work with global fits
seems to indicate that using the PHOJET unfolding, the rate of rise with energy 
of the hadronic cross-section for $\gamma \gamma$ is the same as that for 
$pp, \bar p p$. But this is not quite consistent with the observation by the L3
collaboration that fits  to their data of the form $A s^\epsilon + B s^{-\eta}$ 
gave them values of $\epsilon$ much higher than the $0.09$ quoted for 
the $pp/ p \bar p$.  Secondly the global fit does require a renormalization 
of both
the L3 and OPAL data. Further, this means that effectively one is using  
parton densities for the $\gamma$ quite different than those experimentally
measured. These, as we know, are quite different for the $p$ and the $\gamma$.
Since the $\gamma$ 'owes' part of its hadronic structure to the 'hard'
$q \bar q \gamma$ vertex, should we actually not expect a different behavior 
for the $\gamma$ than for a proton? On the other hand, the EMM model predicts
an uncomfortably fast rise with energy of $\sigma^{tot}_{\gamma \gamma}$
beyond the LEP energy range whereas the BN
model requires a possible breakdown of factorization for a consistent 
description of {\em both} the normalization and energy rise of
all the three, $pp/p\bar p, \gamma p$ and $\gamma \gamma$, cross-sections.
Luckily newer measurements of  $\sigma^{tot}_{\gamma p}$ and 
$\sigma^{tot}_{\gamma \gamma}$ are expected to be available from H1 and OPAL 
soon. We can only look forward to that.

\section{Future Physics Studies of the Photons and with the Photons}
The next generation $e^+e^-$ and $e \gamma$ colliders will offer an 
excellent chance to study the structure of photon, both real and virtual, via
the measurements of hard processes. There have been no new developments 
in that area since the discussions at the PHOTON-01\cite{michael}. 
Measurement  of quantities such as total hadronic cross-sections 
in $ 2 \gamma$ processes will offer possibilities for studying the  
nonperturbative QCD dynamics and model building for calculation of soft
quantities such as cross-sections, as discussed partially in section~\ref{32}. 
In the last two years a very complete study of the physics potentials of the
Compton Collider option was made, using realistic energy and polarization
spectra of photons expected at the Photon Linear Colliders (PLC) after 
inclusion of nonlinear effects as well as using
realistic detector simulations. Various aspects of these studies at 
the future colliders were discussed at the conference~\cite{deroeck}.
I will mention some aspects of the Higgs and CP studies that photons
make possible.

I also review the discussions at the conference of the use of 
the data from low-energy $e^+e^-$ colliders to determine the all important 
hadronic contribution to $(g-2)_\mu$, 
role that $\gamma \gamma$ colliders can play in studying physics beyond the
SM and a method, for helicity amplitude calculations, well suited for 
automation. 

\subsection{Higgs and CP studies at the future $\gamma \gamma$ colliders}
The possibility of very accurate measurement of the $\gamma \gamma$ width of 
the SM Higgs boson, at a PLC, where it is produced as a $s$ channel resonance,
has been discussed at previous photon meetings\cite{Kramer:2000nx}. Since these
original studies, further investigations on accuracy of measurements of the 
$\Gamma (H \rightarrow \gamma \gamma)$ using the $b \bar b, WW$ and $ZZ$ 
channel  as well as getting information on the the phase of 
$\gamma \gamma H$ coupling using the decay of the Higgs in $WW,ZZ$ channel,
with realistic detector simulations and 
careful consideration of higher order corrections and backgrounds, for a 
Higgs with mass $180 < m_H < 350 $ GeV at a PLC, have been performed 
recently\cite{maria_2}.
\begin{figure}[htb]
\includegraphics*[width=17pc]{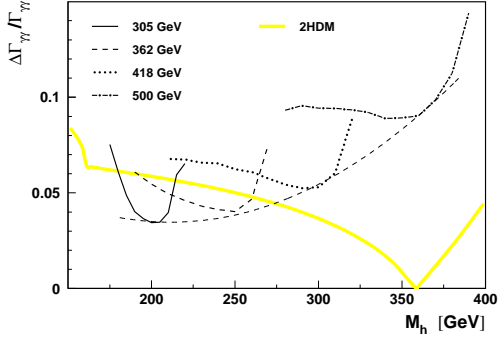}
\caption{ Statistical error on the  measurement of the $\gamma \gamma$ width of
a $H$ as a function of Higgs mass at a PLC\protect\cite{deroeck}}
\label{fig11}
\end{figure}
These show that  a PLC will offer possibilities for discrimination of an 
observed Higgs 
resonance from the SM case. Figure~\ref{fig11}  shows expectations for a 
two Higgs 
Doublet Model (2HDM) along with the possible sensitivity of measurement.
These analyses have also shown that $\Gamma(H \rightarrow \gamma \gamma) 
BR (H \rightarrow b \bar b)$ can be measured to an accuracy varying
between $1.8\%$ to $6.8\%$ as the mass of the Higgs varies from 120 GeV to 160 
GeV, for a SM Higgs.

If the Higgs is heavy enough to decay in to a $t \bar t$ then the CP property
of the Higgs can be determined by using the interference of the $s$ channel
resonance amplitude with the SM QCD production process.  This gives rise to 
{\em mixed} asymmetries in the polarization of the initial laser beam which is
backscattered and the charge of the lepton coming from the $t/ \bar t$ decay.
\begin{figure}[htb]
\includegraphics*[width=17pc]{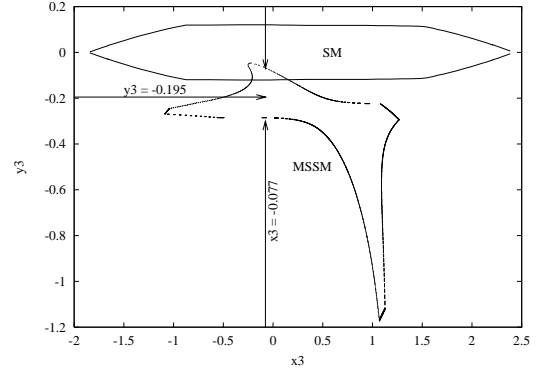}
\caption{Sensitivity of a PLC for separating between a SM and MSSM 
Higgs using  $\gamma \gamma \rightarrow H \rightarrow t \bar t 
\rightarrow lX$\protect\cite{Godbole:2002qu}}
\label{fig12}
\end{figure}
Fig.~\ref{fig12} shows the sensitivity to the CP violating couplings possible 
for a MSSM Heavy Higgs.

The interesting thing is that these  asymmetries can be nonzero even 
for the SM contribution due to the $V-A$ nature of the $tbW$ coupling.
Existence of similar nonzero charge asymmetries which arise from the left
handed nature of the $W$ coupling to fermions was pointed out at the conference
\cite{ginzburg}. They  studied reactions
$\gamma \gamma \to \mu^+\mu^-+\nu\bar\nu$ and $\gamma \gamma \to
W^\pm\mu^\mp +\nu(\bar\nu)$ for $\sqrt{s} > 200$ GeV
and showed that the differences in the distributions of positive
($\mu^+$) and negative charged leptons ($\mu^-$), give rise to observable 
{\em charge asymmetry} of muons which will have nothing to do with any new
physics. In their calculations they take into account all the possible
diagrams and {\em not} just the double resonant ones.
\begin{figure}[htb]
{\centerline{
\includegraphics[height=4cm,width=4cm]{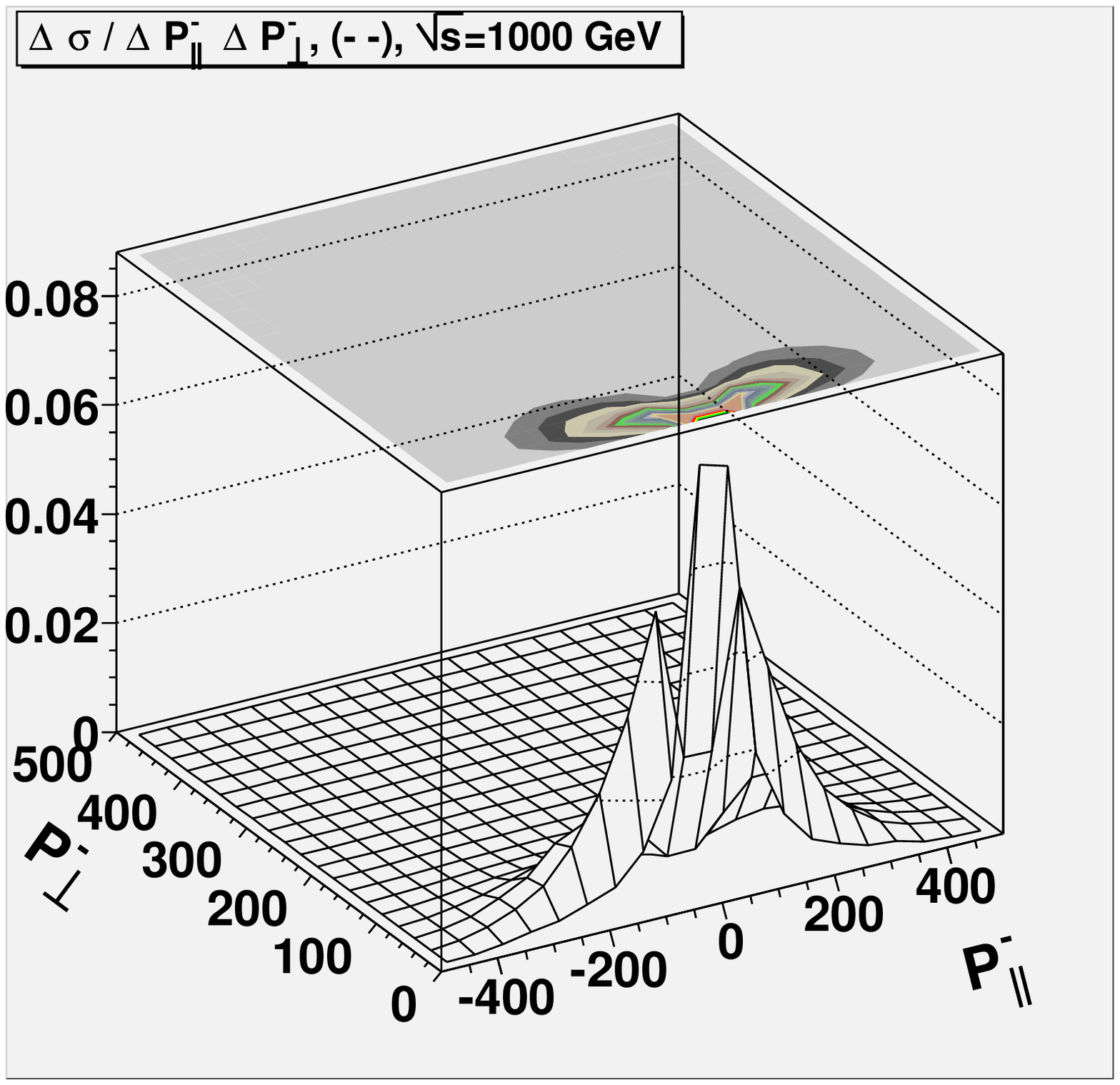}
\includegraphics[height=4cm,width=4cm]{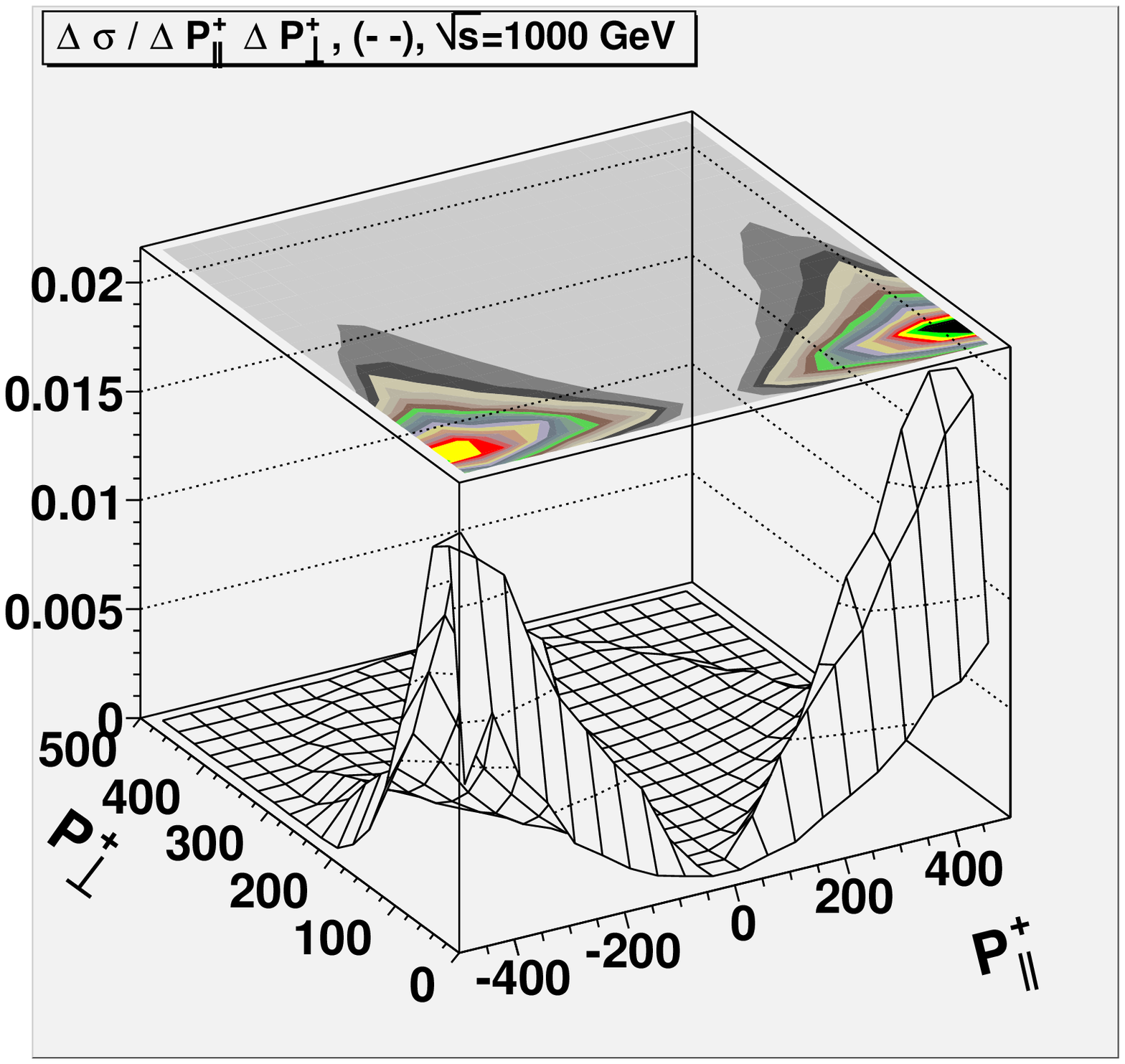}}}
\caption{\sf The distributions in the $(p_\parallel,p_\perp)$
plane for $(-\, -)$ helicity of colliding photons: $\mu^-$ on the
left, $\mu^+$ on the right for $\sqrt{s}=1000$ GeV and 
monochromatic beams\protect\cite{ginzburg}.}
\label{fig13}
\end{figure}
They present the first results on the distributions of
muons $\partial^2 \sigma/(\partial p_{\parallel}\partial
p_{\perp})$ (with $p_{\parallel}$ the component of
$\vec{p}_{\mu^\pm}$ parallel to the collision axis (taken to be
the $z$ axis) and $p_{\perp}=\sqrt{p^2_x+p^2_y}$ the transverse
momentum). Fig.~\ref{fig13} shows such muon distributions in
the $(p_\|,\,p_\bot)$ plane for a $(-\,-)$ initial photon
polarization state with monochromatic photons. We see the difference in
the distributions between the negative and positive muons.

\subsection{Physics beyond the SM  and photons}
Many current investigations of Physics beyond the SM involve modification
of the geometry of our space time; be it extra dimensions, warped or otherwise,
or non-commutative geometry. In all these cases there are interesting effects 
on the photon interactions and hence PLC offers very many possibilities of 
testing/studying  these ideas. Apart from  a summary\cite{deroeck} of the
reach of leptonic and photonic collider  for Extra Dimensional theories, 
the conference also had an experimental 
presentation\cite{kawamoto} of constraints on such theories placed by 
consideration of the LEP data. In one formulation of the non-commutative
QED, the $e e \gamma$ vertex receives a kinematic phase which depends
on the energy momentum tensor $\theta_{\mu \nu}$ as well as $p_\mu$. Thus
Lorentz invariance is violated and there is an unique direction in space. The
modification of the $e e \gamma$ vertex, affects the behavior of all the 
high energy processes involving $\gamma$ such as $e \gamma, e^+ e^-$ and 
$\gamma \gamma$ scattering. The process which could be studied at LEP
was $e^+ e^- \rightarrow \gamma \gamma$, which can probe the scale of 
noncommutavity $\lambda_{NC}$, as these effects modify the polar angle and 
azimuthal angle distribution.
Of course, since there is a unique direction in space the effects now 
will have
a time dependence due to earth's motion. They studied three quantities:
1) ${d \sigma \over d \cos\theta}$ : $\phi$ integrated and time averaged,
2) ${d \sigma \over d \phi}$ : $\cos \theta$ integrated and time averaged
and 3) $\sigma(t) $ : $\phi$ integrated and $\cos \theta$ integrated.
The effect on the $\theta$ distribution is almost independent of $\eta$,
the polar angle of the unique direction, whereas the $\phi$ distribution, 
even after time averaging, has a weak dependence on it. As a result in the end
the non-observation of any deviation from the SM expectations of the process 
$e^+ e^- \rightarrow \gamma \gamma $, puts a lower limit between
$120$ GeV to $180$ GeV, on $\lambda_{NC}$ as a
function of $\eta$. The time dependence gives similar constraints 
as a function of the azimuthal angle $\zeta$.

\subsection{Hadronic contribution to $(g-2)_\mu$}
Evaluation of the  hadronic correction to the $(g-2)_\mu$ has become a very
important subject due to the highly accurate measurements of $(g-2)_\mu$
that have become available recently and the consequent effectiveness of this
quantity as a probe for physics beyond the SM. The precision of the 
experimental measurement will increase further after the full data set 
has been analyzed.  The currently reported deviations of the experimental 
measurement from the  prediction in the SM, are of the same order as 
the error in the theoretical evaluation of the hadronic contribution to 
$(g-2)_\mu$ and hence it is the limiting factor in the accuracy of 
the theoretical prediction of  the SM for the $(g-2)_\mu$
The subject was reviewed at the conference\cite{jegerlehner} and it was 
pointed out that the evaluation of this hadronic contribution uses the 
experimental data on hadron production in $e^+ e^-$ annihilation for 
$\sqrt{s} < 2 $ GeV. Analyses using just the experimental data alone have
an accuracy of $\sim 1.3 \%$. The accuracy of theoretical predictions is 
improved in a data driven analyses which replace data by perturbatively 
calculated R-ratios (the ratio of hadronic cross-section in units of the
$e^+ e^- \rightarrow \mu^+ \mu-$ ). A large theoretical effort is underway
as to how to reduce the theoretical errors in the  evaluation of the
hadronic contribution to $(g-2)_\mu$, by calculation of EW radiative 
corrections to the quantities involved. Further, a much more reliable
evaluation of the vacuum polarization tensor by using the data, is possible
if one uses the Adler function monitored evaluation. 
It is concluded that the necessary accuracy in the prediction, 
so as to become truly sensitive to physics beyond the SM, can be achieved only
if R ratio is measured to better than $1 \%$ up to the $J/\psi$ energies.


\subsection{New calculational techniques for high energy processes}
Accurate theoretical predictions of different  SM processes at the current
and future high energy colliders, require a very high level of computational 
effort  in calculating the required scattering  amplitudes. Hence  numerically
stable and efficient methods for calculating these are always welcome. At this
meeting a new method to calculate jet like QED processes, whose cross-sections
do not drop with increasing energy was discussed\cite{serbo}. 
These typically involve exchange of a virtual photon in $t$ channel. 
The dominant contribution to these cross-sections comes from the small 
scattering angles. The authors are able to write, for arbitrary helicities
of the initial state leptons,  the matrix element in a factorized form
${\cal M}_{fi} = {s \over q^2} J_1 J_2$ where the impact factors $J_1,J_2$ 
are analytical, compact expressions independent of $s$. Thus essentially the 
spinor structures for real or virtual leptons have got replaced by transition
vertices for real leptons. The approximation made in deriving these expressions
omits only terms of the order of ${\theta_i}^2, m_i/E_i \theta_i$ or 
$m_i^2 /E_i^2$, where $\theta_i = P^i_T/E_i$. Since terms of the order 
$m_i/{P_i^T}$ are kept this is a quite different approximation than used in
(say) CALCUL. Further, the factors $J_1,J_2$ are calculated after the
compensating terms are canceled analytically. Hence the evaluation of the 
helicity amplitudes is numerically stable. This method, still to be set
for QCD, certainly holds a lot of promise for the automatic evaluation of 
helicity amplitudes.

\section{Concluding remarks}
It is somewhat unnatural to give a summary of a summary talk. I want to
simply end by saying that this conference has proved that the world of physics
of photons and physics that we can do with the photons is a very alive subject.
Various developments discussed at this conference have implications for a 
lot of issues in QCD, QED and even for the physics beyond the SM. In the coming
years, analyses of the data from HERA-II, Hermes, Compass and $e^+e^-$ 
experiments at Frascati, SLAC, Novosibirsk, Bejiing, KEK etc. will 
provide additional information to answer some of the questions that have been
raised at this conference. Structure of the virtual photon, particle and heavy 
flavour production in $\gamma \gamma$, $\gamma p$ and $\gamma^* p$ reactions
need to be studied at higher orders in QCD, as the data have thrown certain
challenges to the theorists here. These along with the areas of diffractive,
soft as well as forward physics from HERA are the areas to look for further 
developments.  Future of photon physics at future colliders
is very bright indeed, as it is clear that such Compton Colliders, if they are 
realized will not only complement the high energy leptonic colliders to solve
conclusively the physics of the spontaneous symmetry breaking, but will also 
prove an enormously rich QCD laboratory.

\section{Acknowledgment}
It is  a pleasure to thank the organizers and especially Giulia Pancheri
for organizing an excellent physics conference and provide very enjoyable 
atmosphere for stimulating discussions. This work was partially supported
by the Department of Science and Technology, India, under project number 
SP/S2/K-01/2000-II.

\end{document}